\documentclass[preprint,12pt, a4paper]{elsarticle}

\usepackage{amssymb}
\usepackage{color}

\usepackage{lineno}
\usepackage{graphicx}
\usepackage{hyperref}
\usepackage{pbox}
\usepackage{multirow}
\journal{SoftwareX}

\begin{document}
\newcommand{\red}[1]{\textcolor{red}{#1}}

\begin{frontmatter}



\title{Aud{E}xp{C}reator: A {GUI}-Based Matlab Tool for Designing and Creating Auditory Experiments with the {P}sychophysics {T}oolbox}


\author{Duc T.~Nguyen\textsuperscript{a*} and Blair Kaneshiro\textsuperscript{a,b}}

\address{\textsuperscript{a}Center for the Study of Language and Information, Stanford University, Stanford, CA 94305 USA\\
\textsuperscript{b}Center for Computer Research in Music and Acoustics, Stanford University, Stanford, CA 94305 USA\\
\textsuperscript{*}Corresponding email: dtn006@ccrma.stanford.edu}

\begin{abstract}
We present AudExpCreator, a GUI-based Matlab tool for designing and creating auditory experiments.  AudExpCreator allows users to generate auditory experiments that run on Matlab's Psychophysics Toolbox without having to write any code; rather, users simply follow instructions in GUIs to specify desired design parameters. The software comprises five auditory study types, including behavioral studies and integration with EEG and physiological response collection systems. Advanced features permit more complicated experimental designs as well as maintenance and update of previously created experiments. AudExpCreator alleviates programming barriers while providing a free, open-source alternative to commercial experimental design software.
\end{abstract}


\end{frontmatter}

% \linenumbers


\section{Motivation and significance}

Auditory experiments are often designed and constructed by writing one or more computer programs to order and present stimuli, as well as record responses from experimental subjects. The Psychophysics Toolbox (Psychtoolbox)~\cite{brainard1997thePsychophysicsToolbox}, a free and open-source Matlab\footnote{\url{https://www.mathworks.com/}} toolbox, provides a variety of resources for building such experimental programs, and several software contributions have been developed around this toolbox~\cite{cornelissen2002theEyelinkToolbox,schwarzbach2011aSimpleFramework,pelli1997}. Even so, to program experiments that integrate automated stimulus presentation and response acquisition can quickly become overly complex in a script-based Matlab environment. Alternative commercial options for neuroscience, cognitive, and behavioral research include Neurobehavioral Systems Presentation\footnote{\url{http://www.neurobs.com/}} and Psychology Software Tools, Inc.\ ePrime.\footnote{\url{https://www.pstnet.com/eprime.cfm}} While these systems offer greater ease of use through experimenter interfaces, they can be costly and still require basic programming knowledge. 

Matlab, though itself a commercial software product, is often already in use for data analysis. Therefore, extending its utility to the experimental study itself at no added cost may be appealing for many.
% In addition, these types of software can be rigid or overly simple for the complex experiment design sometimes necessary in the research. 
Integrating experimental design, data collection, and data analysis completely within Matlab may also mitigate issues and potential data loss related to exporting and importing data of different formats across disparate software packages. 
% along with potential loss of data in the translation. 
Even so, all of the above options require at least basic knowledge of programming. This fact can be daunting for many, and may act as a technical barrier for prospective researchers wanting to enter the field. 
% And even though acquiring basic programming skills is always a benefit in the long run, it should not become a determining factor that gate keeps potential innovative science from being explored.

With these factors in mind, we were motivated to create AudExpCreator, a tool that enables researchers to design and create a variety of auditory experiments that run on Matlab's Psychophysics Toolbox, without needing to code a single line. Instead, the user specifies the experimental and response collection parameters by means of a Graphical User Interface (GUI), and the software automatically translates these specifications into the necessary Psychophysics Toolbox scripts to run the auditory experiment.

The AudExpCreator tool software package is available for download from the Stanford Digital Repository~\cite{nguyen2017AECSDR},\footnote{\url{https://purl.stanford.edu/qk286xn8411}} and is published under a CC0 license.\footnote{\url{https://creativecommons.org/share-your-work/public-domain/cc0/}} The software package contains the following items: 
\begin{enumerate}
    \item \texttt{AudExpCreator.zip}: A compressed archive of the software tool itself.
    \item \texttt{AudExpCreator\_User\_Manual.pdf}: A User Guide containing extensive documentation of the functionalities and features of the software.
    \item \texttt{stim.zip}: An archive of auditory stimuli for getting started with the software.
    \item \texttt{demosBRS.zip}: The completed auditory experiment demo that we present as an illustrative analysis in this paper (Sec.~\ref{sec:illustrative}). 
\end{enumerate}
The following five study types are currently implemented in AudExpCreator: Behavioral Rating Study, Comparison Behavioral Rating Study, Continuous Behavioral Rating Study, EEG Study, and Neurophysiological Study. The software also contains additional features and add-on tools that enable users to develop more complex auditory experiments, as well as maintain and update any  experiment previously created by AudExpCreator. 

This software package started out as a collection of programming templates using the Psychophysics Toolbox to design the music cognition studies of the Music Engagement Research Initiative at Stanford University.\footnote{\url{https://ccrma.stanford.edu/groups/meri/index.html}} To date, the underlying functionalities of the Continuous Behavioral Rating Study, EEG Study, and Neurophysiological Study types have been used in completed research~\cite{kaneshiro2016dissertation,kaneshiro2016neurophysiologicalAndBehavioral,losorelli2017NMEDTPaper}. As other research groups have expressed interest in using the full collection of templates, the initial goal of releasing this software was simply to aggregate the templates into a toolbox to be published and shared freely with the research community. However, after further consideration of the very real issue of programming as a technical barrier---which was evident even within our own research group---we decided not only to aggregate the templates, but also to develop the AudExpCreator GUI to facilitate creation and design of auditory experiments without the need to code at all. Thus, the now completed AudExpCreator tool makes the functionalities of the Psychophysics Toolbox accessible to researchers from a variety of disciplines, whether or not they have programming experience. Because our research group specializes in music cognition studies, AudExpCreator focuses on auditory stimulus presentation.

We hope that in addition to facilitating audio research, this tool will also ease users into the world of programming and encourage them to learn more about it. AudExpCreator is accompanied by an extensive User Guide, which explains the underlying functionalities of the toolbox. Therefore, should users ever want to begin working directly with the code themselves, the code can be modified in various ways (e.g., to create automated experiments with visual stimuli as well).

\section{Software description}

\subsection{Software Structure}

\begin{figure}[h]
\centering
\fbox{\includegraphics[width=\textwidth, trim=0cm 1cm 0cm 1cm, clip=true]{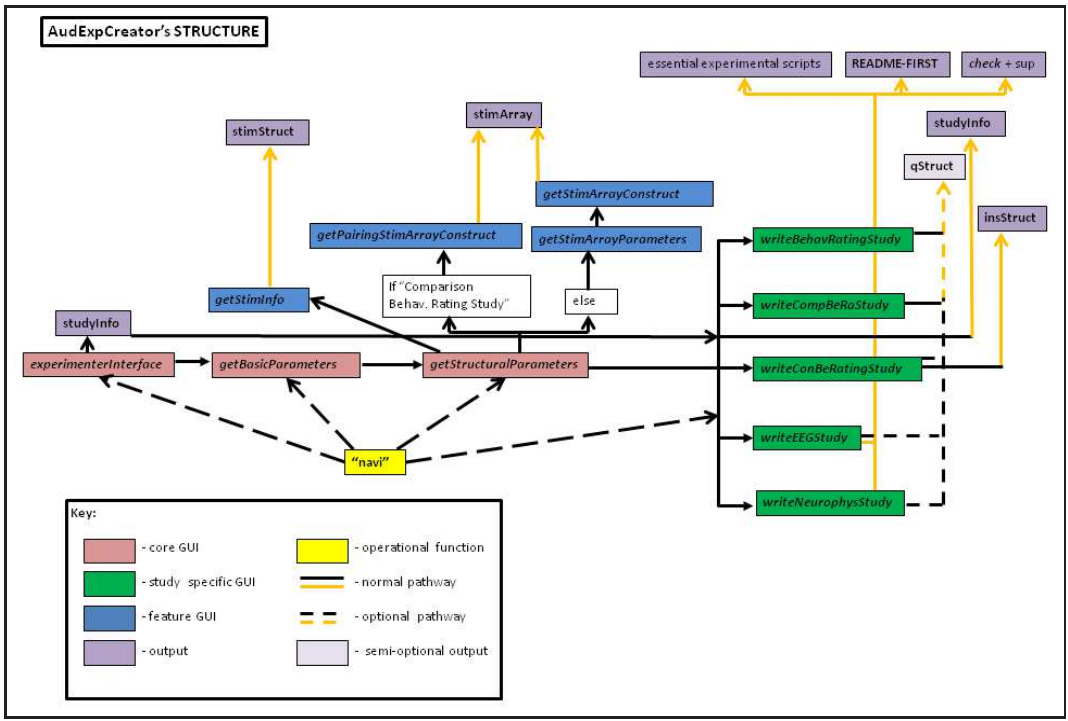}}
\caption{Overall structure of AudExpCreator.}
\label{fig:overview}
\end{figure}

The structure of AudExpCreator (Fig.~\ref{fig:overview}) centers around a set of core GUIs: \textit{experimenterInterface}, \textit{getBasicParameters}, \textit{getStructuralParameters}, and---depending on the type of study users had determined to make---the study-specific GUI. In each of these core GUIs, users are able to gradually design and specify the ``look and feel'' of the auditory experiment, from screen resolution, to background and font color, to the specified stimulus array which defines the presentation ordering of auditory stimuli. Throughout the process of specifying these parameters, AudExpCreator also includes many built-in checks that help ensure that the input parameters and specifications are feasible and will work. Therefore, the only necessary external requirement that the AudExpCreator tool asks of users in the creation of their auditory experiment is the collection of auditory stimuli in .wav format. The process of uploading and organizing these auditory stimuli is facilitated by feature GUIs within the AudExpCreator; however, experimental manipulations of the stimuli themselves are beyond the scope of AudExpCreator and are expected to be completed by users prior to upload. 

Once users have filled out the specifications from the core GUI, finishing with the study-specific GUI, AudExpCreator will generate the auditory experiment. It will create essential experimental scripts, a ``README-FIRST'' file customized for the present experiment, and an additional ``check $+$ sup'' add-on tool that will help users maintain the working integrity of their experiment and assist with small updates to the experiment. 

Additionally, to simplify the navigation of AudExpCreator, the function \textit{navi} was created to help navigate and guide users throughout the process of creating and designing their auditory experiment. Typing ``navi'' into the Matlab console will help them initiate new experimental designs. In addition, should users leave an experimental design unfinished, they can simply return at a later date and type ``navi'' to pick up where they last left off. 
	
To be clear, AudExpCreator does not itself depend on the Psychophysics Toolbox to function. It is completely independent and will function with Matlab version 7.14 (R2012a) or higher. However, the auditory experiments that AudExpCreator generates for users do require the Psychophysics Toolbox in order to run.

\subsection{Software Functionalities}

AudExpCreator contains a variety of features and functionalities. Its primary purpose, however, is to help users generate a fully automated auditory experiment that presents stimuli (with event timing if necessary) plus any questions for subjects to answer; and to collect the subject responses and output those data in .mat format for analysis. The five different types of experimental study types offered by the software package are as follows.

The Behavioral Rating Study is the simplest of the five available studies. Its function is to simply present subjects with auditory stimuli and subsequent behavioral rating questions. The main functions of this experiment type are to automate the following: 1. Presentation of an auditory stimulus or set of stimuli; 2. Presentation of user-specified questions to be answered with behavioral ratings; 3. An interface enabling the subject to answer those questions; 4. Collection and compilation of responses and response times; 5. Looping through steps 1-4 as needed for each stimulus or stimulus set; 6. Termination of the experimental session and data output in .mat format.

The Comparison Behavioral Rating Study focuses on presenting subjects with pairs of auditory stimuli, then allowing the subject to compare between the stimuli in the pair. The main functions of this experiment type are to automate the following: 1. Successive presentation of two auditory stimuli with proper labelling display; 2. Presentation of one or more user-specified comparison questions; 3. An interface enabling the subject to answer those questions; 4. Collection and compilation of responses and response times; 5. Looping through steps 1-4 as needed for all auditory stimulus pairs; 6. Termination of the experimental session and data output in .mat format.

While the first two study types enable retrospective responses, which are delivered after stimuli have finished playing, the Continuous Behavioral Rating Study acquires continuous responses from subjects while an auditory stimulus plays. The main functions of this type of auditory study experiment are to automate the following: 1. Presentation of instructions regarding the user-specified continuous behavioral task; 2. Presentation of one or more auditory stimuli with simultaneous subject interfacing (onscreen analog slider with mouse control) for continuous response acquisition; 3. Acquisition of the continuous response at a sampling rate of approximately 20~Hz; 4. Optional presentation of one or more behavioral rating questions following presentation of auditory stimuli; 5. Compilation of continuous responses and optional behavioral rating responses and response times; 6. Looping through steps 1-5 as needed for each stimulus or stimulus set; 7. Termination of experimental session data output in .mat format.
% as two .mat files for continuous responses data and optional behavioral rating data respectively.  

As part of our own focus on music cognition research, we also work with acquisition of EEG and physiological responses during stimulus presentation. Therefore, the last two studies enable integration of auditory experiments with an external neurophysiogical response acquisition system, specifically the Electrical Geodesics, Inc.\ (EGI) system.

The EEG Study includes all of the integral components of the Behavioral Rating Study while additionally communicating with the EEG acquisition system through two different methods, TTL or TCP/IP, to deliver the triggers (event labels) of auditory stimuli at their onsets. This is so that the EEG data frame is properly annotated with time stamps and triggers. The main functions of this type of experiment are to automate the following: 1. Initiating the communication device or opening the communication channel with the EEG acquisition system; 2. Presentation of auditory stimuli or stimulus sets with simultaneous trigger delivery to the EEG acquisition system; 3. Optional presentation of user-specified behavioral rating questions; 4. Subject interfacing to answer those questions, with responses also sent as triggers to the EEG acquisition system; 5. Optional collection and compilation of responses and response times; 6. Looping through steps 1-5 as needed for each stimulus or stimulus set; 7. Termination of the experimental session and optional output of response data in .mat format.

Finally, the Neurophysiological Study mirrors that of the EEG Study with an additional functionality to accommodate baseline recordings recommended when recording neurophysiological responses such as EKG, EMG, respiratory inductive plethysmography, etc.\ from EGI's Polygraph Input Box (PIB). It allows and mediates the presentation of a baseline auditory stimulus before experimental stimuli, so that readings of neurophysiological responses can be acquired while the subject is at rest. Like the EEG Study, all triggers pertaining to stimulus onsets or behavioral ratings are delivered via TTL or TCP/PI, and additional .mat response data can also be output if desired. Therefore, its main functions are to automate the following: 1. Initiating the communication device or opening the communication channel with the EEG and PIB acquisition system; 2. Presentation of a baseline auditory stimulus before each experimental stimulus or stimulus set with simultaneous onset trigger delivery to the EEG acquisition system; 3. Optional presentation of user-specified behavioral rating questions; 4. Subject interfacing to answer those questions with response triggers also sent to the EEG acquisition system; 5. Optional collection and compilation of responses and response times; 6. Looping through steps 1-5 as needed for each stimulus or stimulus set; 7. Termination of the experimental session and optional output of response data in .mat format.  

Within these five study types, AudExpCreator offers several customization options including stimulus organization, labeling, and grouping; stimulus array construction with shuffling, randomization, probability selection, etc.; basic update and maintenance in the form of an add-on GUI; and finally, advanced updates for extreme overhaul of already built experiments. Detailed information on the five studies and their functionalities, as well as these additional features, can be found in the accompanying User Guide. 

\section{Illustrative Examples}
\label{sec:illustrative}

\subsection{Creating a Behavioral Rating Study}

We will now walk through the process of creating a Behavioral Rating Study experiment, which we will call \textit{demoBRS}. Note that the AudExpCreator source download page includes a file called \texttt{stim.zip}. This file contains the 12 auditory stimuli (.wav format) used in the creation of this demo. These stimuli were obtained from an open data repository~\cite{kaneshiro2012SDR}. The example experiment that is described below is also available on the AudExpCreator source download page, in the file \texttt{demoBRS.zip}. Note that in order to run the completed demo, the Psychophysics Toolbox must be properly installed and the \textit{check} GUI must be run to update the necessary paths. See the User Guide on the download page (\texttt{AudExpCreator\_User\_Manual.pdf}) for more information on the \textit{check} GUI.

\begin{figure}[h]
\centering
\includegraphics[scale=1]{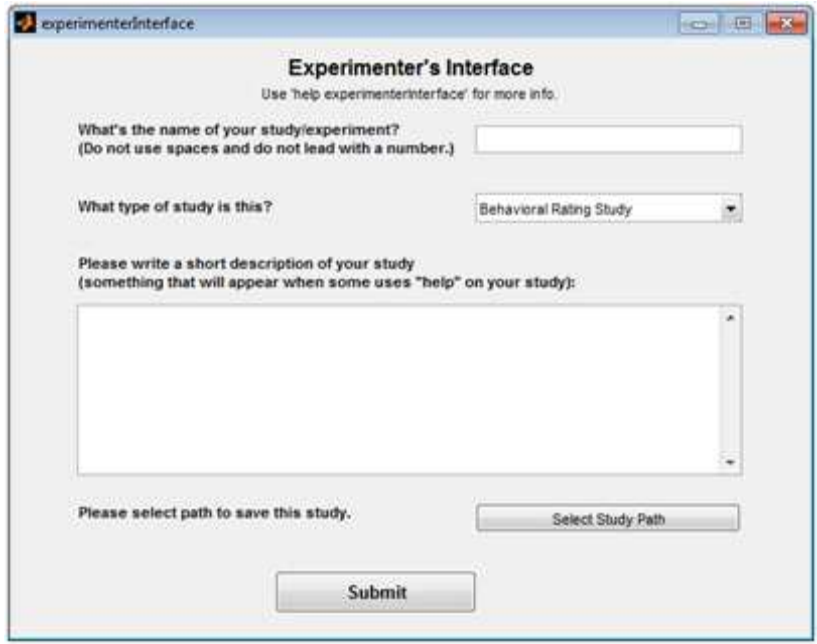}
\caption{experimenterInterface window of the first core GUI.}
\label{fig:gui1ExpInterface}
\end{figure}

First, to begin the process of creating the \textit{demoBRS} auditory experiment, users should open Matlab and change the directory to the AudExpCreator folder. 
% When this is successful users will see the following shown in \textbf{Fig. 2}. 
% \begin{figure}[h]
% \centering
% \includegraphics[scale=1]{InsideAudExpCreator.eps}
% \label{fig:insideFolder}
% \caption{{Inside AudExpCreator Folder}}
% \end{figure}
Next, in the Matlab Command Window, type ``navi'' and, following the instructions, type ``yes'' to allow \textit{navi} to initiate a new study. The AudExpCreator logo will appear, shortly followed by the first core GUI, the experimenterInterface GUI as shown in Fig.~\ref{fig:gui1ExpInterface}. 

\begin{figure}[ht]
\centering
\includegraphics[width=\textwidth]{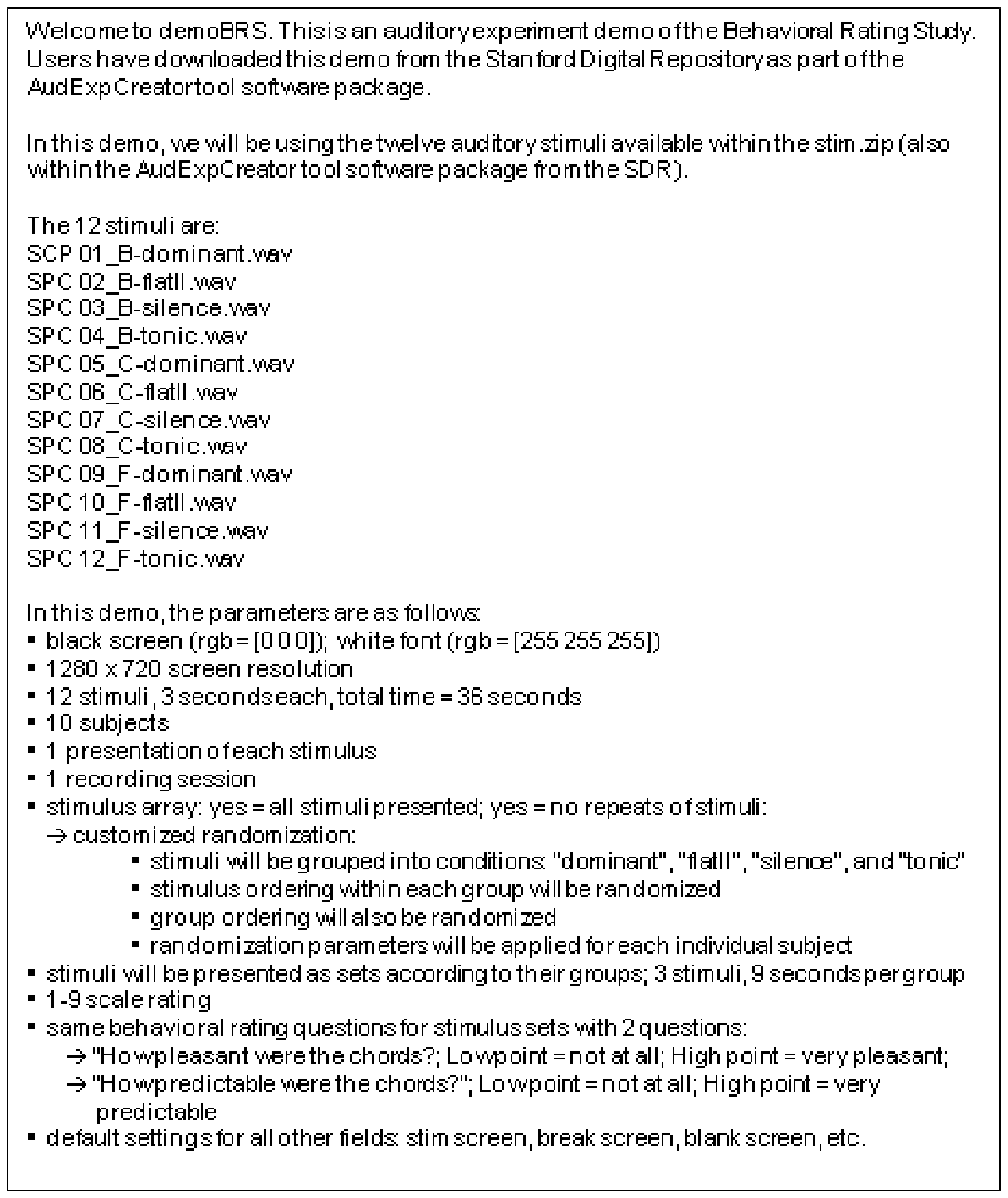}
\caption{Detailed parameters as input to \textit{Study description} of demoBRS.}
\label{fig:demoParameters}
\end{figure}

\begin{figure}[ht]
\centering
\includegraphics[scale=1]{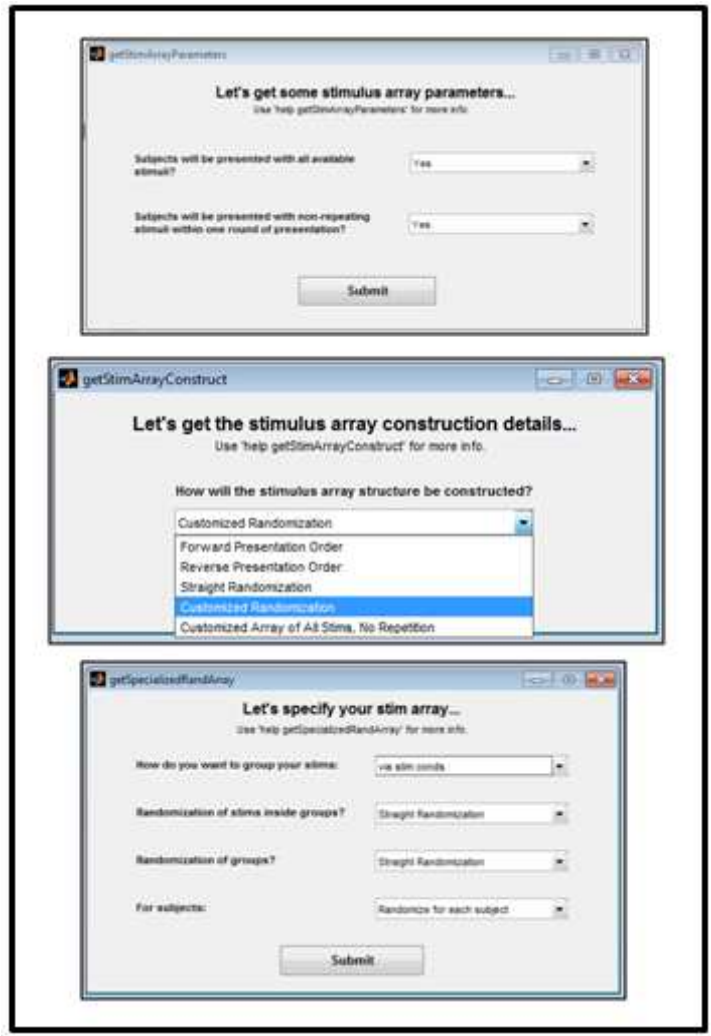}
\caption{{GUI selections for stimulus array creation of demoBRS.}}
\label{fig:stimArraySequence}
\end{figure}

From here, begin by filling in ``demoBRS'' for \textit{Study name}; select ``Behavioral Rating Study'' from the pop-up menu for \textit{Study type}; fill in the \textit{Study description} as shown in  Fig.~\ref{fig:demoParameters}; and supply detailed parameters as shown in Fig.~\ref{fig:demoParameters} for the sequence of GUIs that follows. Because the parameters for this demo call for a customized randomization based on stimulus condition, while filling out the features GUI (\textit{getStimInfo}), users should provide stimulus information as shown in Table~\ref{tab:stimInfo}. In order to create the stimulus array for this customized randomization, users should provide the option selections of feature GUIs as shown in Fig.~\ref{fig:stimArraySequence}.

%\begin{figure}[h]
%\centering
%\includegraphics[scale=1]{stimInfoGUI.eps}
%\caption{{\textbf{\textit{stimInfo GUI}}}}
%\label{fig:stimInfoGUI}
%\end{figure}

% \begin{figure}[h]
% \centering
% \includegraphics[scale=.8]{stiminfo.eps}
% \label{tab:stimulus info}
% \caption{{Table 2. stimulus info (make this a table please)}}
% \end{figure}

\begin{table}[h]
    \centering
    \small
    \begin{tabular}{|l|l|l|l|l|}
    \hline
       \textbf{Stim file}  & \textbf{Stim title} & \textbf{Stim artist} & \textbf{Stim type} & \textbf{Stim condt} \\
       \hline
        \multirow{2}{*}{SCP 01\_B-dominant.wav} & Simple Chord & \multirow{2}{*}{unknown} & \multirow{2}{*}{B Key} & \multirow{2}{*}{dominant} \\ & Progression 01 & & & \\
        \hline
        \multirow{2}{*}{SCP 02\_B-flatII.wav} & Simple Chord & \multirow{2}{*}{unknown} & 
        \multirow{2}{*}{B Key} & \multirow{2}{*}{flatII} \\ & Progression 02 & & & \\
        \hline
        \multirow{2}{*}{SCP 03\_B-silence.wav} & Simple Chord & \multirow{2}{*}{unknown} & \multirow{2}{*}{B Key} & \multirow{2}{*}{silence} \\ & Progression 03 & & & \\
        \hline
        \multirow{2}{*}{SCP 04\_B-tonic.wav} & Simple Chord & \multirow{2}{*}{unknown} & 
        \multirow{2}{*}{B Key} & \multirow{2}{*}{tonic} \\ & Progression 04 & & & \\
        \hline
        \multirow{2}{*}{SCP 05\_C-tonic.wav} & Simple Chord & \multirow{2}{*}{unknown} & 
        \multirow{2}{*}{C Key} & \multirow{2}{*}{dominant} \\ & Progression 05 & & & \\
        \hline
        \multirow{2}{*}{SCP 06\_C-tonic.wav} & Simple Chord & \multirow{2}{*}{unknown} & 
        \multirow{2}{*}{C Key} & \multirow{2}{*}{flatII} \\ & Progression 06 & & & \\
        \hline
        \multirow{2}{*}{SCP 07\_C-tonic.wav} & Simple Chord & \multirow{2}{*}{unknown} & 
        \multirow{2}{*}{C Key} & \multirow{2}{*}{silence} \\ & Progression 07 & & & \\
        \hline
        \multirow{2}{*}{SCP 08\_C-tonic.wav} & Simple Chord & \multirow{2}{*}{unknown} & 
        \multirow{2}{*}{C Key} & \multirow{2}{*}{tonic} \\ & Progression 08 & & & \\
        \hline
        \multirow{2}{*}{SCP 09\_F-dominant.wav} & Simple Chord & \multirow{2}{*}{unknown} & \multirow{2}{*}{F Key} & \multirow{2}{*}{dominant} \\ & Progression 09 & & & \\
        \hline
        \multirow{2}{*}{SCP 10\_F-flatII.wav} & Simple Chord & \multirow{2}{*}{unknown} & 
        \multirow{2}{*}{F Key} & \multirow{2}{*}{flatII} \\ & Progression 10 & & & \\
        \hline
        \multirow{2}{*}{SCP 11\_F-silence.wav} & Simple Chord & \multirow{2}{*}{unknown} & \multirow{2}{*}{F Key} & \multirow{2}{*}{silence} \\ & Progression 11 & & & \\
        \hline
        \multirow{2}{*}{SCP 12\_F-tonic.wav} & Simple Chord & \multirow{2}{*}{unknown} & 
        \multirow{2}{*}{F Key} & \multirow{2}{*}{tonic} \\ & Progression 12 & & & \\
        \hline
        
    \end{tabular}
    \caption{Stimulus information as input to demoBRS.}
    \label{tab:stimInfo}
\end{table}

\begin{figure}[h]
\centering
\includegraphics[scale=.8]{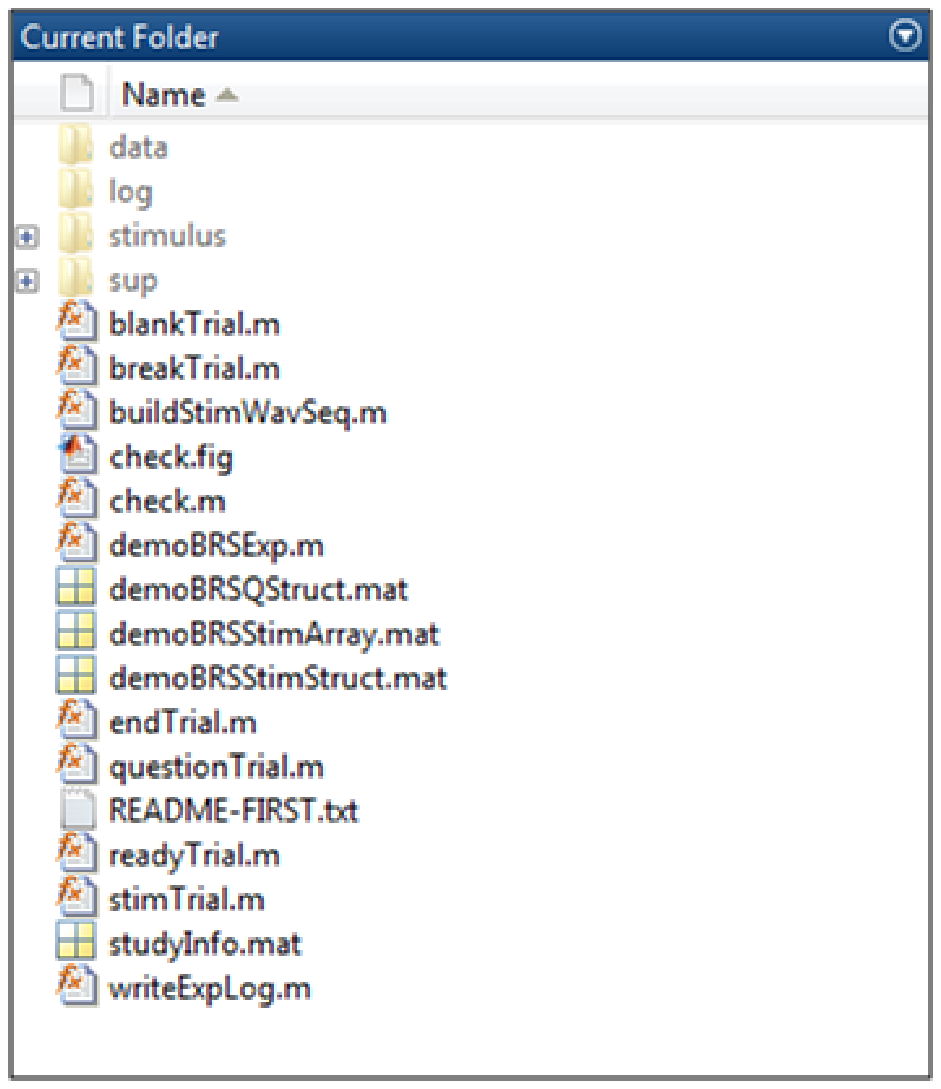}
\caption{{Inside the completed auditory experiment of the demoBRS folder.}}
\label{fig:completedExptFolder}
\end{figure}

\begin{figure}[h]
\centering
\fbox{\includegraphics[scale=.5]{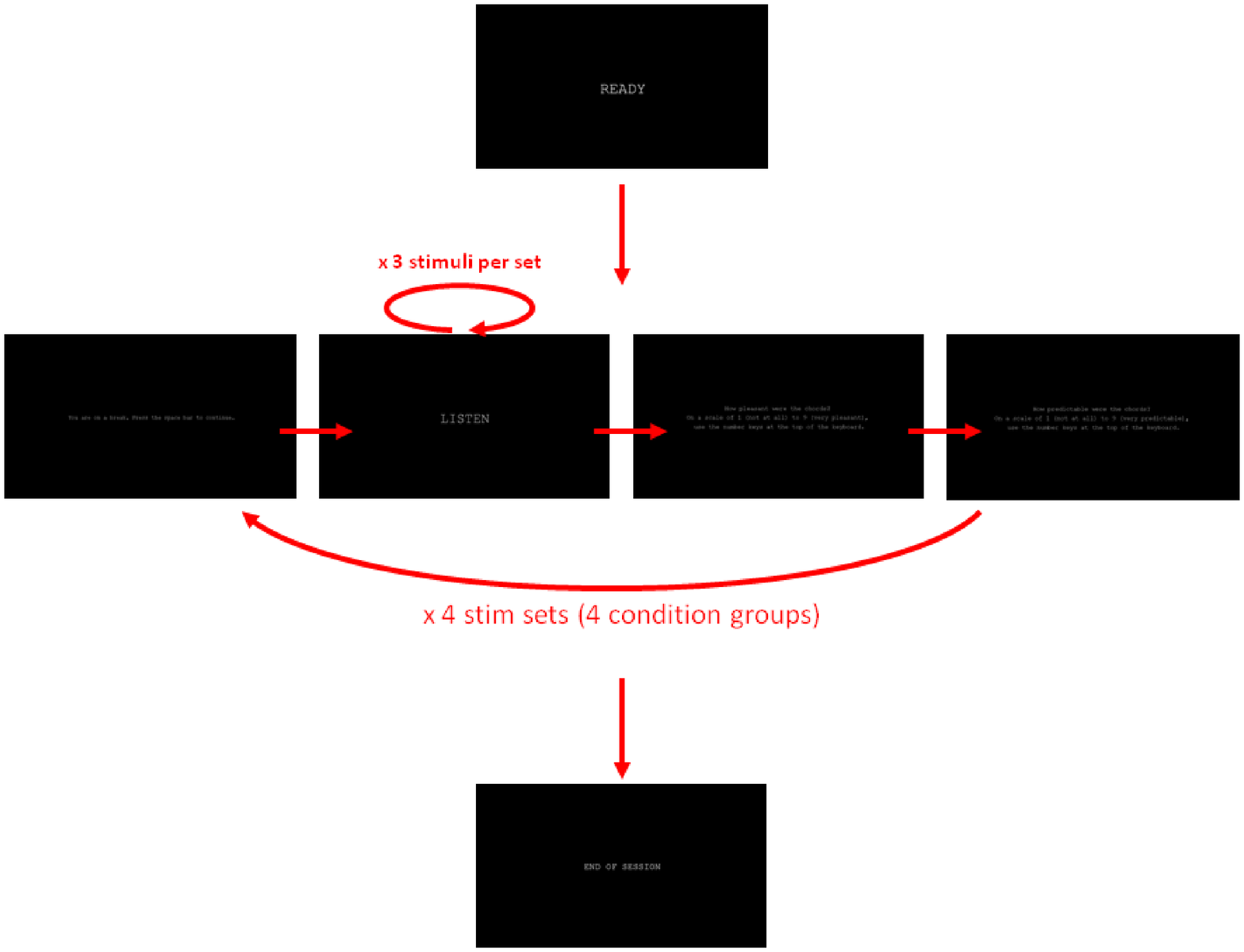}}
\caption{{Presentation sequence overview of the demoBRS auditory experiment.}}
\label{fig:presentationDemo}
\end{figure}

Finally, when AudExpCreator finishes creating the experiment, it will generate the remaining necessary essential folders and functions for the experiment, a personalized ``README-FIRST'' file containing instructions geared toward the parameters of demoBRS, and an additional ``\textit{check} GUI + sup'' add-on tool. Fig.~\ref{fig:completedExptFolder} shows the content in the demoBRS experimental folder necessary to run the demo. This concludes the walkthrough of experiment creation a simple Behavioral Rating Study. To run the study, users should read the ``README-FIRST'' file, as it contains instructions on how to run the study with Matlab. Fig.~\ref{fig:presentationDemo} shows the presentation sequence of demoBR during its run.

\section{Impact and Conclusions}
\label{sec:conclusion}

AudExpCreator is designed to impact the fields of auditory research, facilitating behavioral and neuroscience research in this community. With AudExpCreator the capability to design and create automated auditory experiments is enhanced and greatly simplified. Researchers can quickly generate new auditory experiments in a fraction of the time it takes to program and code experiments themselves. For more advanced programmers, AudExpCreator is a helpful tool expedite experiment creation, leaving more time for heavy-duty data analysis. As AudExpCreator also creates accessible scripts, the expert programmer can use utilize this software simply as a template and subsequently modify the output scripts as needed. 

For researchers with no programming background, AudExpCreator acts as a powerful tool to lower the barrier to entry with programming and working with Matlab. As shown in Sec.~\ref{sec:illustrative}, creating a simple Behavioral Rating Study requires no programming with AudExpCreator. For these researchers, all that is needed to initiate an auditory experiment is to use AudExpCreator following its GUI sequence, and fill out their desired experimental parameters. By simplifying the process, it is hoped that AudExpCreator will encourage new types of scholars in the fields of music cognition, auditory research, and neuroscience (e.g., music theorists, composers, medical practitioners, etc.) to conduct experimental research by making experiment creation---and programming---more accessible, allowing these fields to expand and grow into new types of research endeavors. 

Finally, as AudExpCreator is Matlab based and the auditory experiments it generates themselves use the Psychophysics Toolbox and Matlab, the need for users to acquire additional third-party software for experiment creation is reduced. Consequently, AudExpCreator helps to streamline experimental design, stimulus presentation, data collection, and data analysis completely within Matlab.
% Having everything streamline through Matlab also helps eliminate the issue of data import and export and any potential data loss from those process. 

As previously mentioned, early forms of the AudExpCreator have already been employed to generate various auditory experiments for research studies. At the moment, AudExpCreator offers many functionalities and features and is fully operational in its current form. We plan to continue improving and updating the software as we receive more feedback and feature requests from users. Ultimately, as an open-source tool, we hope that AudExpCreator will be adopted by other researchers and perhaps developed further with additional experimental templates and response acquisition integrations. Though it was created with the fields of music cognition, auditory research, and neuroscience in mind, the future of this tool may find unanticipated utility with researchers from other fields who end up adopting it for their own purposes.  

\section*{Acknowledgements}

This research is supported by the Patrick Suppes Gift Fund (DTN, BK) and the Roberta Bowman Denning Fund for Humanities and Technology (BK). The authors thank Alan Huang for beta testing the software, and members of the Music Engagement Research Initiative at Stanford University for helpful feedback during software development.



%% References:
\section{References}
 \bibliographystyle{elsarticle-num} 
 \bibliography{AudExpCreate}


% \end{thebibliography}

\section*{Required Metadata}

\section*{Current code version}

Ancillary data table required for subversion of the codebase. Kindly replace examples in right column with the correct information about your current code, and leave the left column as it is.

\begin{table}[!h]
\begin{tabular}{|l|p{6.5cm}|p{6.5cm}|}
\hline
\textbf{Nr.} & \textbf{Code metadata description} & \textbf{Please fill in this column} \\
\hline
C1 & Current code version & v1.0 \\
\hline
C2 & Permanent link to code/repository used for this code version & $https://purl.stanford.edu/qk286xn8411$ \\
\hline
C3 & Legal Code License   & CC0 \\
\hline
C4 & Code versioning system used & Stanford Digital Repository \\
\hline
C5 & Software code languages, tools, and services used & Matlab \\
\hline
C6 & Compilation requirements, operating environments \& dependencies & Matlab version 7.14 (R2012a) or later; Psychophysics Toolbox 3 \\
\hline
C7 & If available Link to developer documentation/manual & $https://stacks.stanford.edu/file/druid:qk286xn8411/AudExpCreator_User_Manual.pdf$\\
\hline
C8 & Support email for questions & audExpCreator@gmail.com\\
\hline
\end{tabular}
\caption{Code metadata (mandatory)}
\label{} 
\end{table}

% \section*{Current executable software version}
% \label{}

% Ancillary data table required for sub version of the executable software: (x.1, x.2 etc.) kindly replace examples in right column with the correct information about your executables, and leave the left column as it is.

% \begin{table}[!h]
% \begin{tabular}{|l|p{6.5cm}|p{6.5cm}|}
% \hline
% \textbf{Nr.} & \textbf{(Executable) software metadata description} & \textbf{Please fill in this column} \\
% \hline
% S1 & Current software version & For example 1.1, 2.4 etc. \\
% \hline
% S2 & Permanent link to executables of this version  & For example: $https://github.com/combogenomics/$ $DuctApe/releases/tag/DuctApe-0.16.4$ \\
% \hline
% S3 & Legal Software License & List one of the approved licenses \\
% \hline
% S4 & Computing platforms/Operating Systems & For example Android, BSD, iOS, Linux, OS X, Microsoft Windows, Unix-like , IBM z/OS, distributed/web based etc. \\
% \hline
% S5 & Installation requirements \& dependencies & \\
% \hline
% S6 & If available, link to user manual - if formally published include a reference to the publication in the reference list & For example: $http://mozart.github.io/documentation/$ \\
% \hline
% S7 & Support email for questions & \\
% \hline
% \end{tabular}
% \caption{Software metadata (optional)}
% \label{} 
% \end{table}

\end{document}